# A New Variable Threshold and Dynamic Step Size Based Active Noise Control System for Improving Performance

P. Babu
Department of ECE
K.S. Rangasamy College of Technology
Tiruchengode, Tamilnadu, India.

A. Krishnan
Department of ECE
K.S. Rangasamy College of Technology
Tiruchengode, Tamilnadu, India

*Abstract*— Several approaches have been introduced in literature for active noise control (ANC) systems. Since FxLMS algorithm appears to be the best choice as a controller filter, researchers tend to improve performance of ANC systems by enhancing and modifying this algorithm. In this paper, modification is done in the existing FxLMS algorithm that provides a new structure for improving the tracking performance and convergence rate. The secondary signal y(n) is dynamic thresholded by Wavelet transform to improve tracking. The convergence rate is improved by dynamically varying the step size of the error signal.

## I. INTRODUCTION

Acoustic noise problems become more and more evident as increased numbers of industrial equipment such as engines, blowers, fans, transformers, and compressors are in use. The traditional approach to acoustic noise control uses passive techniques such as enclosures, barriers, and silencers to attenuate the undesired noise [1], [2]. These passive silencers are valued for their high attenuation over a broad frequency range; however, they are relatively large, costly, and ineffective at low frequencies. Mechanical vibration is another related type of noise that commonly creates problems in all areas of transportation and manufacturing, as well as with many household appliances.

Active Noise Control (ANC) [3]–[4] involves an electro acoustic or electromechanical system that cancels specifically, an anti-noise of equal amplitude and the primary (unwanted) noise based on the principle of superposition; opposite phase is generated and combined with the primary noise, thus resulting in the cancellation of both opposite phase is generated and combined with the primary noise, thus resulting in the cancellation of both noises.

The most popular adaptation algorithm used for ANC applications is the FxLMS algorithm, which is a modified version of the LMS algorithm [5]. The schematic diagram for a single-channel feed forward ANC system using the FxLMS algorithm is shown in Fig.1. Here, P (z) is primary acoustic path between the reference noise source and the error microphone and S (z) is the secondary path following the ANC (adaptive) filter W (z). The reference signal x (n) is filtered through S (z), and appears as anti-noise signal y' (n) at the error microphone. This anti-noise signal combines with the primary noise signal d (n) to create a zone of silence in the vicinity of the error microphone. The error microphone measures the residual noise e (n), which is used by W (z) for its adaptation to minimize the sound pressure at error microphone.

Here $\hat{S}(z)$ account for the model of the secondary path S (z) between the output of the controller and the output of the error microphone. The filtering of the reference signals x (n) through the secondary-path model $\hat{S}(z)$ is demanded by the fact that the output y (n) of the adaptive controller w (z) is filtered through the secondary path S (z). [7].

The main idea in this paper is to further increase the performance of FxLMS algorithm in terms of Signal to noise ratio. In modified FxLMS, secondary signal y' (n) is soft threshold dynamically with respect to error signal by wavelet transform to improve the tracking performance. The step size is also varied dynamically with respect to the error signal. Since error at the beginning is large, the step size of the algorithm and the threshold are also large. This in turn increases convergence rate. As the iteration progresses, the error will simultaneously decrease. Finally, the original step size and the threshold will be retained.

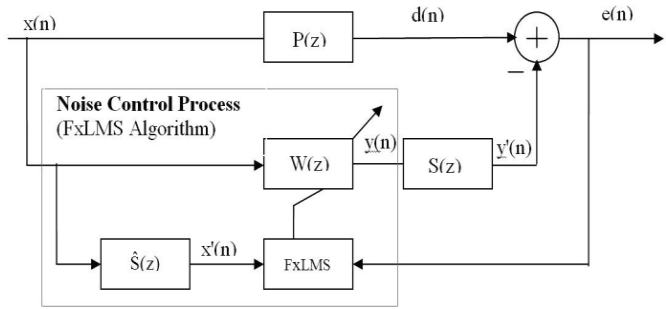

Fig.ure 1. Block diagram of FxLMS based feed forward ANC system.

The organization of this paper is as follows. Section II describes the Secondary path effects. Section III describes FxLMS algorithm. Section IV introduces Wavelet transform. Section V describes the proposed method. Section VI describes the simulation results and Section VII gives the conclusion.

## II. SECONDARY PATH EFFECTS

In ANC system, the primary noise is combined with the output of the adaptive filter. Therefore, it is necessary to compensate $\hat{S}(z)$ for the secondary-path transfer from $y(n)$ to $e(n)$, which includes the digital-to-analog (D/A) converter, reconstruction filter, power amplifier, loudspeaker, acoustic path from loudspeaker to error microphone, error microphone, preamplifier, anti-aliasing filter, and analog-to-digital (A/D) converter. The schematic diagram for a simplified ANC system is shown in Figure 2.

From Fig. 2., the -transform of the error signal is

$$E(z) = [P(z) - S(z)W(z)](X(z)) \quad (1)$$

We shall make the simplifying assumption here that after convergence of the adaptive filter, the residual error is ideally zero [i.e., E (z) =0]. This requires $W(z)$ realizing the optimal transfer function.

$$W^o(z) = \frac{P(z)}{S(z)} \quad (2)$$

In other words, the adaptive filter has to simultaneously Model $P(z)$ and inversely model $S(z)$. A key advantage of this approach is that with a proper model of the plant, the system can respond instantaneously to changes in the input signal caused by changes in the noise sources. However, the performance of an ANC system depends largely upon the transfer function of the secondary path. By introducing an equalizer, a more uniform secondary path frequency response is achieved. In this way, the amount of noise reduction can often be increased significantly [8]. In addition, a sufficiently high-order adaptive FIR filter is required to approximate a rational function $1/S(z)$ shown in "(2)". It is impossible to compensate for the inherent delay due to $S(z)$ if the primary path $P(z)$ does not contain a delay of at least equal length.

## III. FXLMS ALGORITHM

The FxLMS algorithm can be applied to both feedback and feed forward structures. Block diagram of a feed forward FxLMS ANC system of Figure 1. Here P (z) accounts for primary acoustic path between reference noise source and error microphone. $\hat{S}(z)$ is obtained offline and kept fixed during the online operation of ANC. The expression for the residual error e (n) is given as

$$e(n) = d(n) - y'(n) \quad (3)$$

Where y' (n) is the controller output y (n) filtered through the secondary path S (z). y'(n) and y(n) computed as

$$y'(n) = s^T(n)y(n) \quad (4)$$

$$y(n) = w^T(n)x(n) \quad (5)$$

Where w (n) = $[w_0(n)\ w_1(n)\ \ldots\ w_{L-1}(n)]^T$ is tap weight vector, $x(n) = [x(n)\ x(n-1)\ \ldots\ x(n-L+1)]^T$ is the reference signal picked by the reference microphone and s(n) is impulse response of secondary path S(z). It is assumed that there is no acoustic feedback from secondary loudspeaker to reference microphone. The FxLMS update equation for the coefficients of W (z) is given as:

$$w(n+1) = w(n) + \mu e(n)x'(n) \quad (6)$$

Where $x'(n)$ is reference signal x (n) filtered through secondary path model $\hat{S}(z)$

$$x'(n) = \hat{s}^T(n)x(n) \quad (7)$$

For a deep study on feed forward FxLMS algorithm the reader may refer to [7].

## IV. WAVELET THRESHOLDING

The principle under which the wavelet thresholding operates is similar to the subspace concept, which relies on the fact that for many real life signals, a limited number of wavelet coefficients in the lower bands are sufficient to reconstruct a good estimate of the original signal. Usually wavelet coefficients are relatively large compared to other coefficients or to any other signal (especially noise) that has its energy spread over a large number of coefficients. Therefore, by shrinking coefficients smaller than a specific value, called threshold, we can nearly eliminate noise while preserving the important information of the original signal.

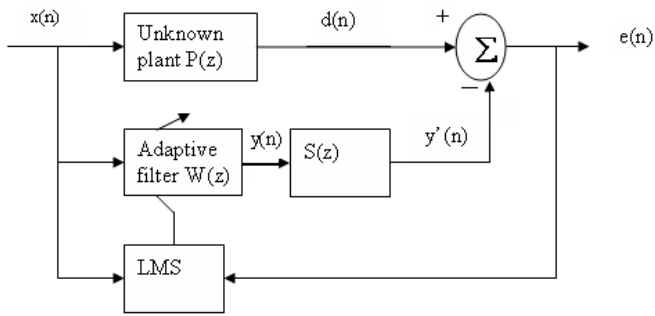

Figure. 2. Block diagram of simplified ANC system

The proposed denoising algorithm is summarized as follow:

i) Compute the discrete wavelet transform for noisy signal.
ii) Based on an algorithm, called thresholding algorithm and a threshold value, shrink some detail wavelet coefficients.
iii) Compute the inverse discrete wavelet transform.

Fig.4 shows the block diagram of the basic wavelet thresholding for signal denoising. Wave shrink, which is the basic method for denoising by wavelet thresholding, shrinks the detail coefficients because these coefficients represent the high frequency components of the signal and it supposes that the most important parts of signal information reside at low frequencies. Therefore, the assumption is that in high frequencies the noise can have a bigger effect than the signal.

Denoising by wavelet is performed by a thresholding algorithm, in which the wavelet coefficients smaller than a specific value, or threshold, will be shrunk or scaled [9] and [10].

The standard thresholding functions used in the wavelet based enhancement systems are hard and soft thresholding functions [11], which we review before introducing a new thresholding algorithm that offers improved performance for signal. In these algorithms, λ is the threshold value and δ is the thresholding algorithm.

A. *Hard thresholding algorithm*

Hard thresholding is similar to setting the components of the noise subspace to zero. The hard threshold algorithm is defined as

$$\delta_\lambda^H = \begin{cases} 0 & |y| \leq \lambda \\ y & |y| > \lambda \end{cases} \qquad (8)$$

In this hard thresholding algorithm, the wavelet coefficients less than the threshold λ will are replaced with zero which is represented in Fig. 3-(a).

B. *Soft thresholding algorithm*

In soft thresholding, the thresholding algorithm is defined as follow :( see Figure 3-(b)).

$$\delta_\lambda^S = \begin{cases} 0 & |y| \leq \lambda \\ \operatorname{sign}(y)(|y|-\lambda) & |y| > \lambda \end{cases} \qquad (9)$$

Soft thresholding goes one step further and decreases the magnitude of the remaining coefficients by the threshold value. Hard thresholding maintains the scale of the signal but introduces ringing and artifacts after reconstruction due to a discontinuity in the wavelet coefficients. Soft thresholding eliminates this discontinuity resulting in smoother signals but slightly decreases the magnitude of the reconstructed signal.

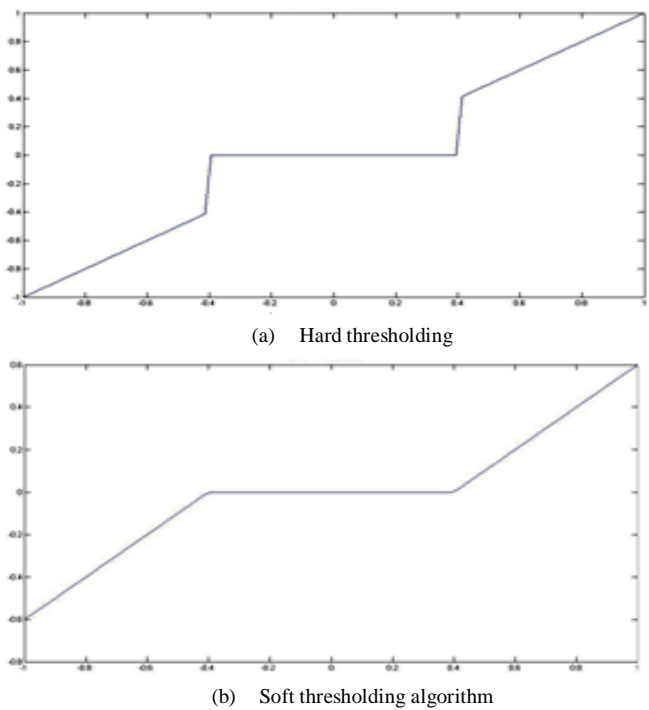

(a)  Hard thresholding

(b)  Soft thresholding algorithm

Figure.3. Thresholding algorithms (a) Hard. (b) Soft

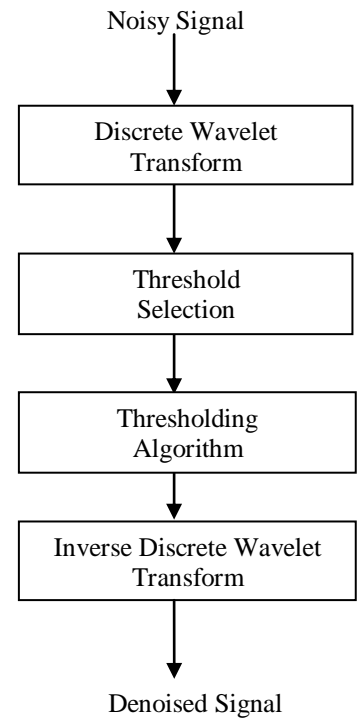

Figure 4. Denoising by wavelet thresholding block diagram

## V. PROPOSED METHOD

### A. Variable thresholding algorithm

In the proposed method $y'(n)$, the secondary signal of FxLMS is denoised by wavelet. This is performed by a thresholding algorithm, in which the wavelet coefficients smaller than a specific value or threshold, will be shrunk or scaled. The signal $y'(n)$ can be soft thresholded because this eliminates the discontinuity and results in smoother signal, such that λ is the threshold value and δ is the thresholding algorithm in order to improving the tracking performance of FxLMS algorithm.

The wavelet transform using fixed thresholding algorithm for signal $y'(n)$ is defined as follow:

$$\delta_\lambda^s = \begin{cases} 0 & |s^T y| \le \lambda \\ \text{sign}(s^T y)(|s^T y| - \lambda) & |s^T y| > \lambda \end{cases} \quad (10)$$

The wavelet transform using fixed soft thresholding will improve the tracking property when compared with traditional FxLMS algorithm based on active noise control systems. The threshold value used in fixed soft thresholding algorithm is $\lambda = 0.45$, since the amplitude of the noise signal is small.

The performance of the system can be further increased by using variable threshold function rather than the fixed threshold function based on the error signal e (n), which is

$$\lambda'' = \frac{\lambda''}{1 - \text{abs}(e(n))} \quad (11)$$

It has been noted that initially the error of the system is large allowing large threshold value λ. As the number of iteration continues, the error of system will decrease. Finally, it retains the original threshold value. The soft thresholding algorithm using variable threshold value is given by below:

$$\delta_\lambda^s = \begin{cases} 0 & |y'| \le \lambda'' \\ \text{sign}(y')(|y'| - \lambda'') & |y'| > \lambda'' \end{cases} \quad (12)$$

Where $y' = s^T y$ is the secondary path signal given in "(4)"

### B. Variable Step Size algorithm

The step size of the FxLMS algorithm is varied dynamically with respect to the error signal. Since error at the beginning is large, the step size of the algorithm is also large. This in turn increases convergence rate. As the iteration progress, the error will simultaneously decrease. Finally, the original step size will be retained.

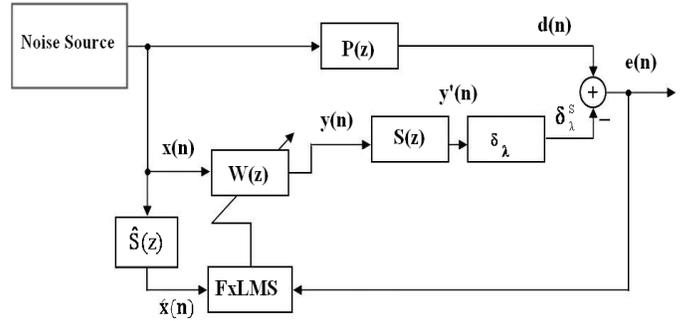

Figure5. Block diagram for proposed method

Fig.5 shows the block diagram for proposed method. Thus the convergence rate of the FxLMS algorithm is improved by varying the step-size as well as wavelet threshold value with respect to error signal. From the Fig. 5, the expression for the residual error e(n) is given as

$$e(n) = d(n) - s^T y \quad (13)$$

Initially the error in the system is very high. So very large step size is selected. Hence the convergence rate is also very high. Then the step size is varied for the instant and the previous value of the error signal e (n). Finally the error is reduced greatly by the implementation of the dynamic step size algorithm.

This idea of dynamic step size and dynamic threshold calculation is represented in "(11)" and "(15)".

$$w(n+1) = w(n) + \mu(n)e(n)x'(n) \quad (14)$$

Where,

$$\mu(n) = \frac{\mu(n)}{1 - abs(e(n))} \quad (15)$$

Thus the "(11')" and "(15)" is called as modified FxLMS algorithm for improving the performance of existing algorithm.

## VI. SIMULATION RESULTS

In this section the performance of the proposed modified FxLMS algorithm with wavelet thresholding is demonstrated using computer simulation. The performance of the variable wavelet thresholding algorithm is compared with fixed wavelet thresholding algorithm on the basis of noise reduction R (dB) and convergence rate is given in "(16)" and "(17)".

$$R\ (dB) = -10 \log\left(\frac{\sum e^2(n)}{\sum d^2(n)}\right) \quad (16)$$

$$\text{Convergence Rate} = 20 \log 10\{abs(g)\} \quad (17)$$

The large positive value of R indicates that more noise reduction is achieved at the error microphone. The computer simulation for modified FxLMS algorithm performance is illustrated in Fig.6. and Fig.7. Fig.6 shows the characteristics of Noise reduction versus number of iteration times. It has been seen that the modified FxLMS with variable soft thresholding and dynamic step-size produce better noise reduction compared with modified FxLMS with fixed soft thresholding.

Fig.7. shows the characteristics of convergence rate in dB with respect to number of iterations. It has been seen that the convergence rate of modified FxLMS with variable soft thresholding and dynamic step-size increases by reducing the number of iterations compared with modified FxLMS with fixed soft thresholding.

Fig.8. shows the characteristics of residual error with respect to number of iterations. It has been seen that the residual error of modified FxLMS with variable soft thresholding and dynamic step-size increases by reducing the number of iterations compared with modified FxLMS with fixed soft thesholding.

Fig.9. shows the characteristics of signal value with respect to number of iterations. Fig.10. shows that the characteristics of signal value with respect to number of iterations. It has been seen that the signal value of modified FxLMS with variable soft thresholding and dynamic step size increases by reducing the number of iterations compared with modified FxLMS with fixed soft threshodling

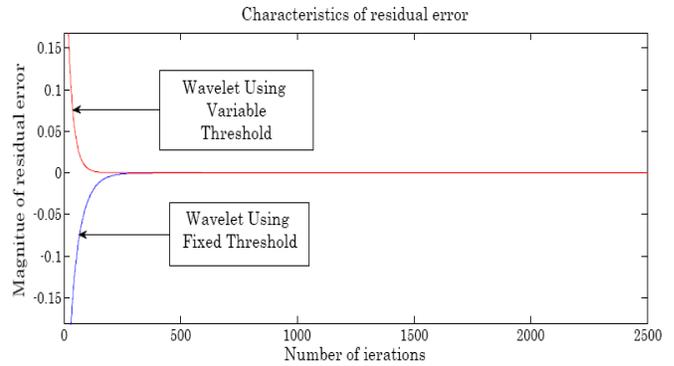

Figure 8. Residual error versus iteration time *(n)*

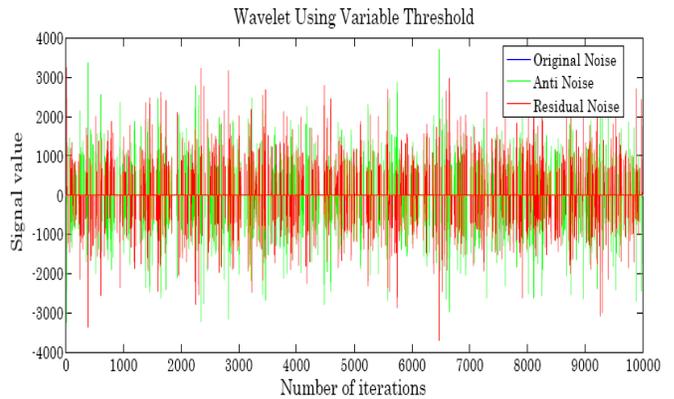

Figure 9. Signal value versus iteration time *(n)*

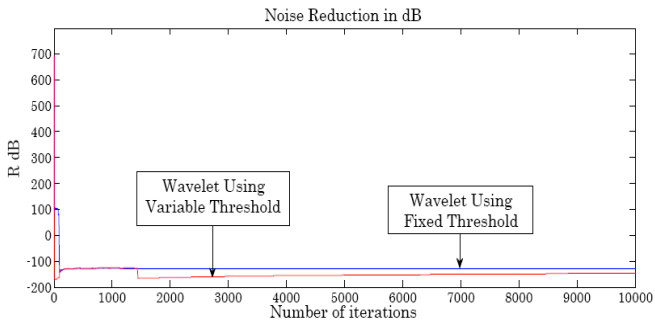

Figure 6. Noise reduction versus iteration time *(n)*

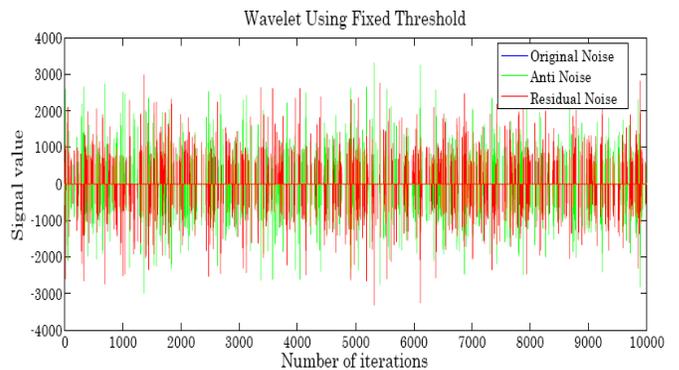

Figure 10. Signal value versus iteration time *(n)*

## VII. CONCLUSIONS

Here we propose a modified FxLMS structure for ANC system. This structure combines the concept of wavelet dynamic soft thresholding with the dynamic variable step size. It shows better tracking performance and convergence rate than the conventional FxLMS algorithm and FxLMS wavelet soft threshold algorithm. The main feature of this method is that it can achieve improved performance than the existing methods.

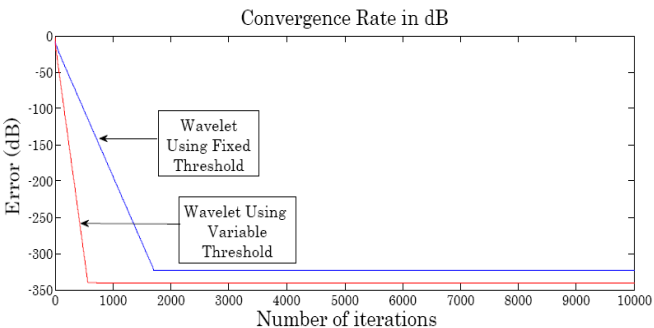

Figure 7. Characteristics of convergence rate

## ACKNOWLEDGMENTS

The authors would like to thank the reviewers for their many insightful comments and useful suggestions. The authors also would like to express their gratitude to our beloved chairman Lion Dr.K.S.Rangasamy and our principal Dr.K.Thyagarajah for supporting this research.

## AUTHORS PROFILE

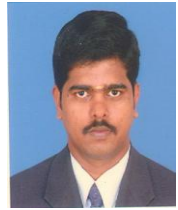

**Babu Palanisamy** received the B.E degree from Madras University, Chennai, India in 1998, and M.E. degree from Madurai Kamaraj University, Madurai, India in 2002. From 2002 to 2007, he worked as a faculty in K.S.Rangasamy College of Technology, Tamilnadu, India. He is currently a Ph.D. candidate in Anna University, Chennai, India. He is a member of IETE and ISTE. His research interests include Signal Processing and Communication Systems.

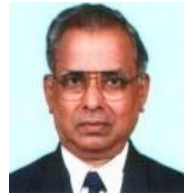

**A.Krishnan** received the Ph. D. degree from Indian Institute of Technology Kanpur, Kanpur, India. He is currently a professor with K. S. Rangasamy College of Technology, Tiruchengode, and Tamilnadu, India. He is a member of IEEE, IETE, and ISTE. His research interests include quality of service of high speed networks and signal processing.